\documentclass[fleqn,usenatbib]
{mnras}

\usepackage{newtxtext,newtxmath}

\usepackage[T1]{fontenc}
\usepackage{ae,aecompl}

\usepackage{graphicx}	
\usepackage{amsmath}	
\usepackage{amssymb}	

\title[Graphene Oxide Nanoparticles in the Interstellar Medium]{Graphene Oxide Nanoparticles in the Interstellar Medium}

\author[P. J. Sarre]{P. J. Sarre,$^{1}$\thanks{E-mail: peter.sarre@nottingham.ac.uk}
\\
$^{1}$School of Chemistry, The University of Nottingham, University Park, Nottingham NG7 2RD, United Kingdom
}

\date{Accepted XXX. Received YYY; in original form ZZZ}

\pubyear{2019}

\begin{document}
\label{firstpage}
\pagerange{\pageref{firstpage}--\pageref{lastpage}}
\maketitle


\begin{abstract}
Dust particles play a major role in the formation, evolution and chemistry of interstellar clouds, stars and planetary systems. Commonly identified forms include amorphous and crystalline carbon-rich particles and silicates. Also present in many astrophysical environments are polycyclic aromatic hydrocarbons (PAHs), detected through their infrared emission, and which are essentially small flakes of graphene. Astronomical observations over the past four decades have revealed a widespread unassigned `Extended Red Emission' (ERE) feature which is attributed to luminescence of dust grains.  Numerous potential carriers for ERE have been proposed but none has gained general acceptance. In this Letter it is shown that there is a strong similarity between laboratory optical emission spectra of graphene oxide and ERE, leading to this proposal that emission from graphene oxide nanoparticles is the origin of ERE and that these are a significant component of interstellar dust. The proposal is supported by infrared emission features detected by the  \textit{Infrared Space Observatory (ISO)} and the \textit{Spitzer Space Telescope}. 
\end{abstract}




\section{Introduction}
\label{sec:Introduction}

\subsection{Background}
\label{subsec:Background}
Extended Red Emission (ERE) is a broad emission feature noted over forty years ago by \citet{coh75} in a spectrophotometric study of the Red Rectangle -- a mixed-chemistry object which comprises a binary star, HD~44179, an oxygen-rich circumstellar disk,  and an extended biconical carbon-rich nebula from which ERE emanates. It has since been detected in a very wide range of Galactic and extragalactic sources \citep{wit14,lai17}. These include reflection nebulae, planetary nebulae, novae, \ion{H}{ii} regions, high-latitude galactic cirrus clouds, the diffuse galactic medium and external galaxies - see \citet{wit88, wit90, sco94, szo98, smi02, rhe07,ber08, wit08}.  ERE varies in peak wavelength (600-850~nm) and width (60-120~nm) both within and between objects.

Numerous carriers for ERE have been proposed but none of these is widely accepted. Proposals and carrier models published up to 2017 are discussed in \citet{lai17}; these include hydrogenated amorphous carbon (HAC), polycyclic aromatic hydrocarbon (PAH) molecules, quenched carbon composite (QCC), C$_{60}$, carbon clusters, silicon and silicate nanoparticles, biofluorescence,  magnesium silicate, nanodiamonds, PAH di-cations, and dimer cations, with hydrogen recently added \citep{Holmlid2018}.  ERE-like luminescence has also been detected in interplanetary dust particles and carbonaceous chondrites \citep{all87,wop88,qui05}.  

In this Letter oxidised graphene nanoparticles, small graphene oxide (GO) grains, are proposed as the ERE carrier based on comparison of laboratory photoluminescence (PL) of graphene oxide with ERE, and similarity in the spatial distribution of ERE and IR emission from chemical groups expected to be present in astronomical GO.

\subsection{ERE Carrier Characteristics}
\label{subsec:Required carrier characteristics}
A wide range of molecules and materials give rise to luminescence at red/near-IR wavelengths and, given the wavelength variation and breadth of the ERE feature, assignment of the carrier on these data alone presents a challenge. However, detailed studies have produced some general constraints \citep{wit04,wit14,lai17}. ERE is observed from carbon-rich but not oxygen-rich nebulae (\citealt{Furton1990}; \citealt{Furton1992}), strongly suggesting that the element carbon is involved; ERE is also seen in mixed-chemistry carbon-oxygen sources such as the Red Rectangle.  Characteristics to be satisfied by proposed carriers include: an ERE peak wavelength -- width relationship and a lack of ERE when the exciting star has T$_{eff}$ $<$ 10,000~K as described by \citet{dar99}, a lower limit on the estimated photon (not energy) conversion efficiency of about 10\% for excitation in the visible-to-far-UV range \citep{gor98}, and a requirement for photons with wavelengths short of 118~nm for ERE excitation in NGC~7023 \citep{wit06}. It is likely significant that a ubiquitous feature of many ERE-emitting sources is aromatic infrared band (AIB) emission from PAHs which is also induced by UV/vis excitation - see reviews by \cite{tie08,tie13}. However, the spatial distribution of ERE- and PAH-emitting regions is not identical from which it is inferred that there is not a direct one-to-one chemical correspondence between ERE carriers and PAHs in general \citep{wit14}. \cite{lai17} have discussed likely photophysical mechanisms giving rise to the ERE emission.

\section{GO nanoparticles and ERE}


Pristine graphene is a single-layer planar network of six-membered carbon rings in an sp$^2$ bonding framework.  It does not possess an optical band gap, the first electronic transition lying in the far-ultraviolet arising from a \begin{math}\pi^*-\pi\end{math} transition. However, exposure of graphene to an oxygen plasma leads to incorporation of oxygen, forming graphene oxide which does have an optical band gap.  \citet{gok09} have shown that laser irradiation of GO generated in this way results in photoluminescence (PL) in the visible spectral range (Figure 1) where spectra for three spatially separate positions on the GO sample are shown with peak wavelengths of 630-660~nm. The spectra for Pos~1 and 2 on the oxidised graphene surface are almost identical and for Pos~3 the PL was bleached intentionally by using intense laser irradiation \citep{gok09}. The PL was assigned as due to CO-related localised electronic states at the oxidation sites \citep{gok09}. In an alternative approach yielding broadly similar results, \citet{luo09} started with as-synthesised GO \citep{luo_2_09} and explored laser-excited PL spectra of solid and aqueous GO as the samples were progressively reduced with hydrazine. 

\begin{figure}
\centering
\includegraphics[scale=1,angle=0.0]{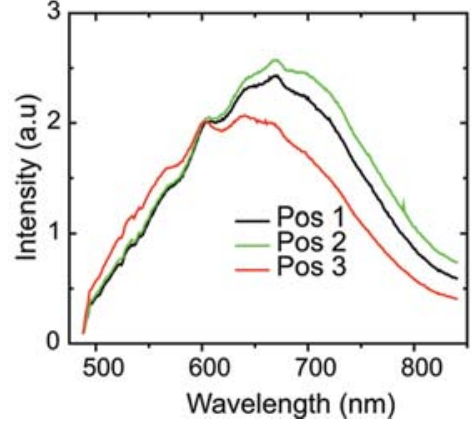}
\caption{PL spectra of GO produced by an oxygen plasma treatment of graphene and excited by laser radiation at 473~nm. Spectra for three irradiated positions (Pos 1, 2 and 3) on the GO sample are shown.  Reprinted with permission from \citet{gok09}. Copyright 2009 American Chemical Society.}
\end{figure}

Figure 2 shows ERE in the Red Rectangle 6$''$ and 10$''$ South of the central star HD~44179 \citep{wit90}.  ERE exhibits variation in peak wavelength and width both within the Red Rectangle, and also between objects; this is illustrated for a wider set of targets in fig. 7 of \citet{led01}. The peak wavelength and FWHM for the laboratory PL of GO (figure 1) and the Red Rectangle ERE (figure 2) at 6$''$ are very similar, indicating that GO is a good candidate as the ERE carrier. 

Laboratory samples of GO are generally produced by exfoliation of graphite oxide using methods such as the modified Hummers method in which four main oxygen-bearing sites are found -- carbonyl (>C=O), epoxide (--O--), carboxylic acid (--COOH) and hydroxyl (--OH) groups, not all of which may be of astronomical relevance. Sites for these groups are illustrated schematically in figure 3.  The chemical structure of GO is a matter of ongoing discussion with a number of models proposed - for a summary see \cite{dre10,zha15}.  In experiments by \citet{cuo11} thermal reduction at $700^{\circ}$C resulted in removal of many oxygen-containing groups illustrated in figure 3, but with carbonyl groups being persistent.  \cite{li12} have found that GO samples with high levels of carbonyl groups produce PL with the best spectral match to ERE. A second luminescence feature of GO is commonly seen near 400~nm. This `blue luminescence' is due to transitions involving confined sp$^2$ bonded `aromatic' regions \citep{chi12,yua19}; the possible astrophysical relevance of this will be discussed elsewhere. 

An important question, addressed in the following section, is whether there is evidence from astronomical observations for IR emission at wavelengths consistent with IR laboratory absorption spectra of GO due to these chemical groups, and whether the spatial distributions of ERE and oxygen-related IR emission features in astronomical objects are similar. 

\begin{figure}
\centering
\includegraphics[scale=0.754,angle=180.5]{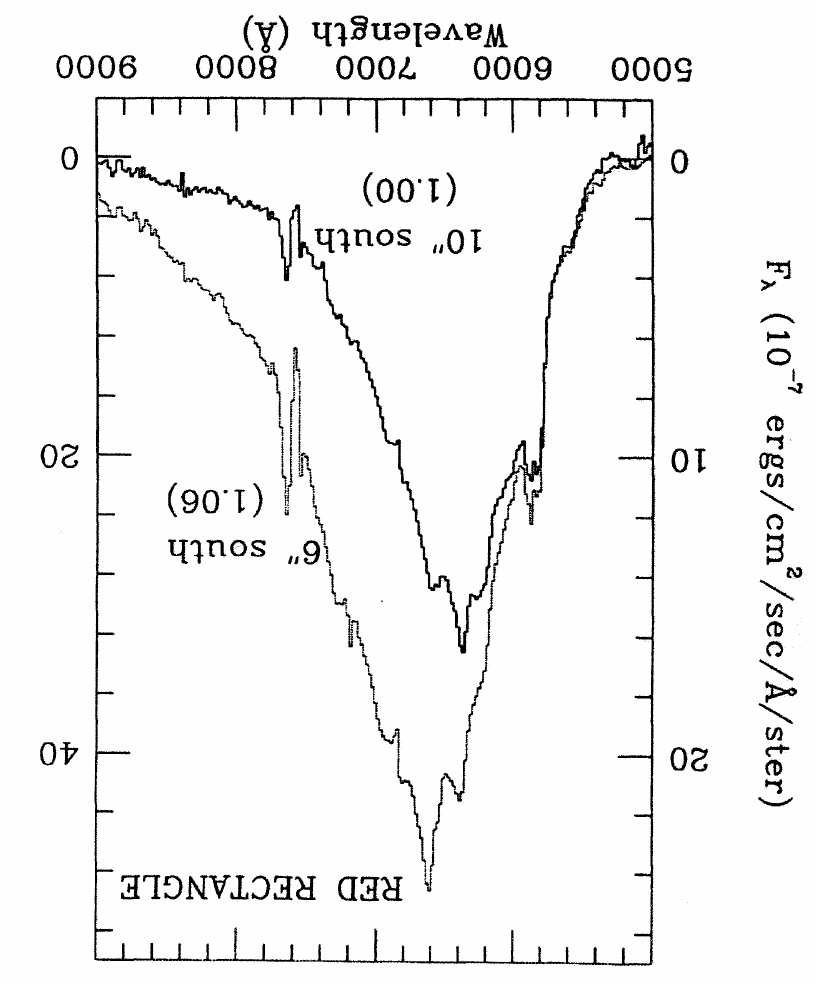}
\caption{ERE in the Red Rectangle nebula 6$''$ and 10$''$ south of HD~44179.  The figures in parentheses are the ratio of ERE to scattered light. Reproduced with permission from \citet{wit90}. The wavelength dispersion in figs 1 and 2 on the page is the same.} 
\label{fig2}
\end{figure}

\begin{figure}
\centering
\includegraphics{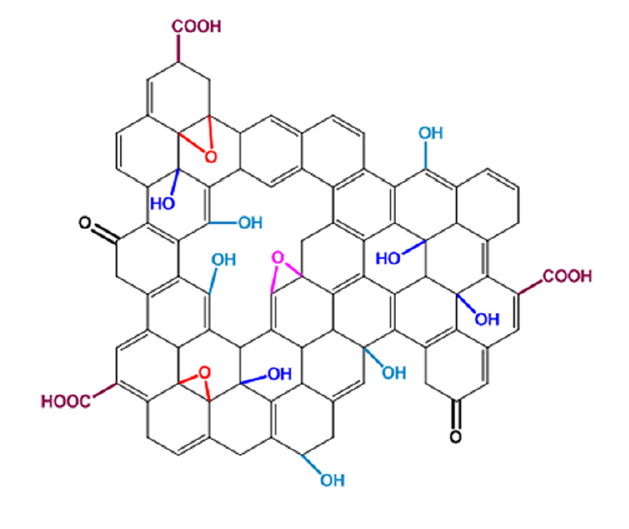}
\caption{Schematic illustration of GO showing sites of oxygen-containing groups in as-produced (modified Hummers method) GO \citep{cuo11}. Hydrogen atoms (C-H) are omitted for clarity. The representation is based on theoretical work by \citet{gao10} and is reprinted from \citet{cuo11}, with the permission of AIP Publishing.} 
\label{fig3}
\end{figure}

\section{Infrared signatures}

 The carbonyl C=O stretch transition is seen in laboratory absorption spectra of GO with a low level of oxidation \citep{kri13}. \textit{ISO} spectra of the Red Rectangle have an emission feature at 6.0$\,\umu$m which is considered to be likely due to a C=O stretch of a quinone-type PAH \citep{pet02}. The ISO spectra are for a 14$''$ x 20$''$ exposure so direct comparison with the established bipolar-shaped ERE distribution is unfortunately not possible. \citet{hsi16} have collated data for 20 sources which exhibit the 6.0$\,\umu$m feature, eight of which are classified as mixed-chemistry objects.  An alternative origin in olefinic double-bond functional groups was suggested \citep{hsi16}.  However, taking gas-phase absorption data from \citet{wal19}, carbonyl-containing molecules with one~(1) or more (2, 3) aromatic rings - p-benzoquinone~(1), 1,4 napthaquinone~(2), anthrone~(3) and anthraquinone~(3) have expected peak emission wavelengths at 6.01, 5.99, 5.99 and 5.98$\,\umu$m, where the commonly invoked 15~cm$^{-1}$ redward shift from absorption to astronomical emission has been invoked as discussed by \citet{bau09}.  From this it is deduced that attribution of the 6.0$\,\umu$m astrophysical emission band to the C=O stretch of oxygenated PAHs is supported by laboratory data.  

In \textit{ISO} spectra of a number of sources, \cite{pet02} found that the 6.0$\,\umu$m band appears not to be correlated with the nearby carbon-carbon stretching band at 6.2$\,\umu$m and this is confirmed by \textit{Spitzer} data for NGC~2023 where the G6.0\footnote{G refers to a Gaussian decomposition} and 6.2  distributions do not match \citep{pet17}. Within the SL FOV for the Southerly \textit{Spitzer }observations of NGC~2023 \citep{pet17}, the G6.0 feature peaks along the S ridge which differs greatly from the 6.2, 7.7 and 8.6$\,\umu$m AIBs, and while it shares in part the distribution of the 11.2$\,\umu$m feature due to neutral PAHs, its distribution is closest to the 8$\,\umu$m `bump' as presented by \citet{pet17}. The G6.0 distribution also has much in common with the 5-10$\,\umu$m and 10-15$\,\umu$m plots, the 14.7$\,\umu$m continuum associated with Very Small Grains (VSGs), and with molecular hydrogen S(2) and S(3) emission as seen earlier in studies of 1-0 S(1) and other highly excited H$_2$ lines which depend on UV excitation \citep{mcc99}. Taking the PAH and continuum \textit{Spitzer} data across NGC~2023, \citet{pet17} comment that the 6.0$\,\umu$m PAH emission `seems to be somewhat unique', and a `unique' spatial distribution is also found for the Northern map.  Within the SL FOV, the S ridge is the region in which ERE emission is strongest \citep{wit89,pil10}, indicating a link between 6.0$\,\umu$m emission and ERE. The SL northern map \citep{pet17} is an area of weak ERE emission \citep{wit89} and a 6.0$\,\umu$m feature is found only in a very small region. 

A second broader band centred at 8.0$\,\umu$m is seen in laboratory spectra of GO and is attributed to C-O-C epoxy group(s) \citep{kri13}. Although not classified as a formal PAH band in the decomposition of NGC~2023 spectra by \cite{pet17} (their figure 2) a fairly broad feature in NGC~2023 is seen and termed the 8$\,\umu$m `bump'. This `bump' is not the same as the wide 8.0$\,\umu$m \textit{Spitzer} IRAC filter which covers the 6.2, 7.7 and 8.6$\,\umu$m PAHs.  Rather the `bump' resembles a PAH-type band, though broader.  Given that the spatial distribution of the `bump' is very similar to the G6.0 band, a related origin for these two bands is suggested.  In future work it would be of  interest to compare the G6.0 and 8.0$\,\umu$m `bump' intensities with ERE strength along sections in NGC~2023, NGC~7023 and other extended objects.  Laboratory IR absorption by hydroxyl groups of GO has been reported near 2.9$\,\umu$m \citep{bag10} but it is much broader than other features and might be difficult to detect if present in the proposed astronomical GO; there is no evidence for such an emission feature in \textit{ISO} spectra of the Red Rectangle. In summary it is found that IR emission from the special sub-group of oxygenated PAH structures (GO) supports the attribution of ERE to GO.  

\cite{pil10} has shown that along sections taken orthogonal to ridges in NGC~2023 and NGC~7023, ERE peaks at the interface of neutral PAH and Very Small Grain (VSG) emission where VSGs with typical size of about 500 carbon atoms are being evaporated to form PAHs - as discussed by \citet{ces00}, \citet{ber07} and \citet{cro16}. This suggests that the GO particle size falls roughly in the same range as the well-studied PAHs.  In the context of this proposal VSGs could include multilayer graphite oxide grains which split into individual GO layers on becoming separated by exposure to UV, shocks etc.

\section{Further comments}

Laboratory studies show that luminescence of GO occurs when it is excited with visible light \citep{gok09}.  This might be considered inconsistent with the astrophysical constraint that UV irradiation is a prerequisite for ERE emission \citep{smi02,wit14}. However, if the formation of GO in astrophysical environments results from decomposition of multi-layered graphite oxide under UV irradiation, the need for UV would be satisfied.  A second possibility is that GO formation depends on creation of radical sites \textit{e.g.} through photo-removal of H from PAH structures, and replacement by O to form carbonyl or epoxy groups. The principle of a two-step process has been invoked for other proposed ERE carriers \citep{wit06,rhe07}. 

The attribution of ERE to GO can be tested by laboratory experiments on GO nanoparticles in which the degree of oxidation is controlled, the temperature of the sample varied and broad-band excitation is employed. In the Red Rectangle the peak wavelength is reported to change from $\sim$ 755~nm about 2$''$ South of HD~44179 \citep{rou95,led01} to $\sim$ 670~nm 6$''$ South and 645~nm 10$''$ South \citep{wit90}. While this might be due to a compositional change in the carrier, it is important to determine the effect of temperature and particle size on GO nanoparticle emission.

If confirmed, the presence of GO nanoparticles could have implications in other areas of astrochemistry. In the Red Rectangle a set of unidentified molecular emission features \citep{sch80}, sits atop the ERE feature, the carriers of which have been suggested to be decomposition products of the ERE-emitting material \citep{sch91}. Could oxygen-containing PAHs and related large carbonaceous molecules be the carriers of these Red Rectangle bands which are very strong near 5,800 \AA\ and 6,600 \AA? These would be large organic `dye' molecules in a chemical sense \citep{sar06}. An association of these emission bands with a subset of diffuse interstellar band carriers has been suggested \citep{sar91,fos91,sar95}. 

Other areas worthy of examination include (i) a known laboratory emission feature of GO near 400~nm which originates in sp$^2$ graphitic islands and which might contribute to astronomical `Blue Luminescence' \citep{vij05}, (ii) any contribution of GO to, and variation in, the 2175 \AA\ interstellar absorption feature and (iii) the `missing oxygen' problem. Catalytic reactions of atoms and molecules on GO surfaces could also be investigated.

\section{Acknowledgement}
I thank the Leverhulme Trust for award of a Leverhulme Emeritus Fellowship.




\bibliographystyle{mnras}
\bibliography{ERErev1bibtex}

%

\bsp	
\label{lastpage}
\end{document}